\documentclass[twocolumn,secnumarabic,amssymb, nobibnotes, aps, prb, superscriptaddress]{revtex4-2}
\usepackage{lipsum}
\usepackage{titlesec}
\titlespacing\section{0pt}{12pt plus 4pt minus 4pt}{4pt plus 20pt minus 2pt}
\usepackage{xcolor}
\usepackage{braket}
\usepackage{amsmath}
\usepackage{comment}
\usepackage{physics}
\usepackage{afterpage}
\usepackage{placeins}
\usepackage{graphicx}
\usepackage{float}
\usepackage{booktabs}
\usepackage{multirow}
\usepackage{array}
\usepackage{setspace}
\graphicspath{{Figs/}}
\usepackage{siunitx}
\usepackage{hhline}
\usepackage{xfrac}
\usepackage{float,graphicx}
\usepackage{mathtools}
\usepackage{listings}
\usepackage{amssymb}
\usepackage{titlesec}
\usepackage{amsfonts}
\usepackage[version=4]{mhchem}

\usepackage{epstopdf}
\catcode`@11
\def\seceqaa{\@addtoreset{equation}{section}
\def\theequation{A\arabic{equation}}}
\def\seceqbb{\@addtoreset{equation}{section}
\def\theequation{B\arabic{equation}}}
\def\seceqcc{\@addtoreset{equation}{section}
\def\theequation{C\arabic{equation}}}
\def\seceqdd{\@addtoreset{equation}{section}
\def\theequation{D\arabic{equation}}}
\def\seceqee{\@addtoreset{equation}{section}
\def\theequation{E\arabic{equation}}}
\def\seceqff{\@addtoreset{equation}{section}
\def\theequation{F\arabic{equation}}}
\def\seceqgg{\@addtoreset{equation}{section}
\def\theequation{G\arabic{equation}}}
\def\seceqhh{\@addtoreset{equation}{section}
\def\theequation{H\arabic{equation}}}
\catcode`@11

\begin{document}

%\date{today}

\title{Colossal anomalous Hall and Nernst effect from the breaking of nodal-line symmetry in Cu$_2$CoSn Weyl semimetal: A first-principles study}

\author{Gaurav K. Shukla}
\affiliation{School of Materials Science and Technology, Indian Institute of Technology (Banaras Hindu University), Varanasi 221005, India}
\author{Ujjawal Modanwal}
%\affiliation{School of Materials Science and Technology, Indian Institute of Technology (Banaras Hindu University), Varanasi 221005, India}
\author{Sanjay Singh*}
\affiliation{School of Materials Science and Technology, Indian Institute of Technology (Banaras Hindu University), Varanasi 221005, India}

%\date{today}

\begin{abstract}
  The presence of topological band crossings near the Fermi energy is essential for the realization of large anomalous transport properties in the materials. The topological semimetals (TSMs) host such properties owing to their unique topological band structure such as Weyl points or nodal lines (NLs), that are protected by certain symmetries of the crystal. When the NLs break out in the system due to perturbation in Hamiltonian, a large Berry curvature arises in the surrounding area of the gapped NL. In the present work, we studied anomalous transport properties of Cu$_2$CoSn compound, which has cubic Heusler crystal structure (space group: Fm$\bar{3}$m). The  Cu$_2$CoSn full Heusler compound possesses three NLs in the absence of spin-orbit coupling close to the Fermi level. These NLs gap out with the consideration of the SOC and a large Berry curvature observed along the gapped NLs. The integral of Berry curvature gives the intrinsic anomalous Hall conductivity (AHC) about 1003 \textit{S/cm} and the anomalous Nernst conductivity (ANC) about 3.98 \textit{A/m-K} at the Fermi level. These values of AHC and ANC are comparable to the largest reported values for the Co$_2$MnGa Heusler compound. Therefore, Cu$_2$CoSn becomes a newborn member of the family of full Heusler compounds, which possesses giant  AHC and ANC that can be useful for the spintronics application.
\end{abstract}

\maketitle
\section{INTRODUCTION}
%\vspace*{-3mm}
The discovery of the Dirac fermions in the topological insulators became a hotspot of research of the past decade in condensed matter physics \cite{wang2017quantum,hasan2021weyl,RevModPhys.81.109,RevModPhys.82.3045}. In recent years, the discovery of the Weyl semimetals  (WSMs) and the related high-fold fermions materials have simulated immense research attention in the topological phase of materials \cite{wang2017quantum,hasan2021weyl}. The WSM is a subset of the Dirac semimetal, where a pair of Weyl points forms due to the breaking of inversion and/or time-reversal symmetry (TRS), which lifts the four-fold degeneracy of the Dirac point \cite{vafek2014dirac,weyl1968gesammelte,burkov2016topological,hasan2017discovery}. WSMs show a variety of interesting phenomena such as chiral anomaly, chiral magnetotransport and anomalous transport response owing to their unique band topology \cite{RevModPhys.90.015001,yan2017topological}. The WSMs due to the breaking of the inversion symmetry (IS) have been discovered widely \cite{xu2015discovery,yang2015weyl,lv2015observation,PhysRevLett.117.146403,xu2015experimental}, while the WSMs result from the breaking of TRS symmetry called magnetic WSMs discovered recently \cite{CTS,prb1}. The advantage of the magnetic WSM over the conventional WSM is that the band topology of magnetic WSMs can be easily tuned via manipulating the magnetic moment direction \cite{CTS,Weyl}. Besides the zero dimension crossing of bands in the WSMs, the higher dimension crossing is also possible, where the bands cross each other along a closed curve called nodal lines (NLs) \cite{burkov2011topological,PhysRevB.90.115111}. These NLs generally protected by the certain symmetry of the crystal. E.g., the TRS and IS can protect the NLs in the absence of spin-orbit coupling (SOC) \cite{PhysRevB.92.081201,kim}. The mirror symmetries with opposite eigenvalues can also protect the NLs both in the presence and absence of SOC \cite{bian2016topological,schoop2016dirac, PhysRevLett.117.016602}.

Anomalous Hall effect (AHE) is a fundamental transport property, which describes the large transverse voltage drop in a current carrying ferromagnetic material even in the zero external magnetic field \cite{nagaosa2006anomalous,nagaosa2010anomalous,tian2009proper,yue2017towards,manna2018heusler,sakuraba2020giant}. AHE got an immense interest in the condensed matter physics for its possible application in spintronic, Hall sensors and as a fundamental tool to detect the magnetization in a small volume, where the magnetometry measurements are not compatible \cite{sensor,ning2020ultra,ohno2000electric}. The AHE arises due to the extrinsic mechanism related to the scattering events as well as the intrinsic mechanism related to the Berry curvature of Bloch bands \cite{smit1955spontaneous,smit1958spontaneous,karplus1954hall, karplus1954hall,sundaram1999wave,xiao2010berry}. The Berry curvature is equivalent to the intrinsic pseudo-magnetic field in the reciprocal space which leads to the transverse deflection of spin-polarized moving charge carriers and develops the intrinsic AHE \cite{xiao2010berry}. 

Anomalous Nernst effect (ANE); another interesting phenomenon that is a counterpart of AHE describes the generation of transverse voltage drop in the material with broken TRS, when subjected to a longitudinal temperature gradient \cite{ikhlas2017large,guin2019anomalous,asaba2021colossal}. The ANE is closely analogous to the AHE \textit{i.e.}  ANE also arises from intrinsic and extrinsic contributions \cite{guin2019anomalous, mizuguchi2019energy}. Several experimental, as well as theoretical studies on ANE, have been reported on magnetic materials \cite{guin2019anomalous,chen2022large,sakai2020iron,PhysRevMaterials.4.024202,guo2017large}.
WSMs are prominent materials for the large AHE and ANE as the Weyl points in the momentum space act as the magnetic monopole and are the source and drain of the Berry curvature \cite{manna2018heusler,guin2019anomalous}. 
Besides the Weyl points, if the  NLs present in the \textit{k}-space gap out due to SOC, the Berry curvature introduces along the gapped NLs and creates the transverse voltage in the system \cite{guin2019anomalous,manna2018heusler}. If the Weyl points or gapped NLs are near the Fermi level their signatures can be observed in the anomalous transport properties of materials  \cite{prb1,liu2018giant}. For \textit{e.g.}, the first discovered magnetic WSMs Co$_3$Sn$_2$S$_2$ shows the large anomalous Hall conductivity (AHC) due to the gapped NLs and the Weyl points present in the system \cite{liu2018giant}. The ANE in the Co$_3$Sn$_2$S$_2$, Mn$_3$X  (X = Ge, Sn) and  Fe$_3$X  (X = Ga, Al) are interesting due to their characteristic low-energy electronics structure including Weyl points near to the Fermi energy \cite{chen2022large,sakai2020iron,PhysRevMaterials.4.024202,guo2017large}. Among the different classes of materials, Heusler alloys are promising for their wide range of properties \cite{graf2011simple,felser2015basics,felser2015heusler}. Recently, Heusler compound attracted much interest as quantum material because some of them are discovered as magnetic WSM due to the co-existence of the magnetism and the topology \cite{prb1,guin2019anomalous,CTS, chang2016room,Weyl}. Heusler compounds also promise the large AHE and ANE due to large  Berry curvature associated with their topological band structure \cite{guin2019anomalous,li2020giant}. The magnetic Heusler compounds also offer the possibility to tune the band topology via manipulating the magnetic moment direction and hence the AHE and ANE can be easily tuned by changing the magnetic moment \cite{CTS}. The largest AHC ($\sim$ 1260 \textit{S/cm} \cite{guin2019anomalous} and 2000 \textit{S/cm} at 2T \cite{sakai2018giant}) and anomalous Nernst conductivity (ANC) ($\sim$ 4 \textit{A/m-k} \cite{sakai2018giant}) so far, reported in the Co$_2$MnGa magnetic Heusler compound.

Cu$_2$CoSn Heusler compound has been identified as the topological semimetal in the topological material database and expected to exhibit large AHC \cite{bradlyn2017topological,vergniory2019complete,ji2022spin}.  
In the present manuscript, we theoretically investigated the structural, magnetic, and anomalous transport properties \textit{i.e.} AHE and ANE in the Cu$_2$CoSn Heusler compound. Cu$_2$CoSn is the ferromagnetic material, which exhibits three NLs in the absence of SOC due to the presence of the three relevant mirror reflection symmetries of the lattice. We found that by switching on the SOC the NLs gap out according to the magnetization direction and a strong Berry curvature originates along the gapped NL, which leads to the large Berry curvature in the system. The Berry curvature calculation gives the  AHC and ANC around $\sim$1000 \textit{S/cm} and $\sim$ 3.98  \textit{A/m-K} at the Fermi level, which is comparable to the largest reported AHC and ANC in the well known Co$_2$MnGa Heusler compound \cite{guin2019anomalous}.
\section{COMPUTATIONAL DETAIL}
 The \textit{ab initio} calculation for the electronic band structure of  Cu$_2$CoSn was performed by employing the density functional theory using the Quantum Espresso code \cite{giannozzi2009quantum}. The Plane wave basis set and the Optimized norm-conserving Vanderbilt pseudo-potentials \cite{PhysRevB.88.085117} were used for the calculation. The plane wave cutoff energy was chosen 80 Ry and the exchange-correlation functional was chosen in the generalized gradient approximation \cite{perdew1996generalized}. The integration in \textit{k}-space was carried out with 8$\times$8$\times$8 grid and the convergence criterion of total energy was chosen 10$^{-8}$ eV. The relaxed lattice parameter was used in the calculation. We extracted the Wannier functions from the DFT bands by Wannier90 code \cite{marzari1997maximally,souza2001maximally}. The maximally localized Wannier functions (MLWFs) for s orbitals on Sn and d orbitals on Cu and Co have been used as the basis of the tight-binding Hamiltonian.  Wanniertool software was used to investigate the topological properties such as NLs and Berry curvature in the two dimensions (2D) reciprocal plane. The Kubo formula implemented in Wannier90 code was used for the calculation of the Berry curvature, which can be given as \cite{Gradhand_2012} 
 \begin{eqnarray}
\Omega^n_{ij} = i \sum_{n \neq n'} \frac{{\langle n|\frac{\partial H}{\partial R^i}|n'\rangle} {\langle n'|\frac{\partial H}{\partial R^j}|n \rangle}-(i\xleftrightarrow{}j)}{(E_n - E_n')^2}
\end{eqnarray}
 Here E$_n$ and $\ket{n}$ are the eigenvalue and eigenstate of the Hamiltonian H. 
 
 The AHC can be calculated using equation;
\begin{equation}
\sigma_{ij} = -{\frac{e^2}{\hbar} \sum_{n}\int\frac{d^{3}\textit{k}}{(2\pi)^3}\Omega^n_{ij}f_n}.  
\end{equation}
Here, f$_n$ represents the Fermi distribution function.

The expression for ANC can be given as \cite{guin2019anomalous};
\begin{multline}
\alpha^A_{ij} (T, \mu) = -\frac{1}{T}\frac{e}{\hbar} \sum_{n}\int\frac{d^{3}\textit{k}}{(2\pi)^3}\Omega^n_{ij}[(E_n-{\mu})f_n +\\
 K_BT\, ln(1+exp(-\frac{E_n-{\mu}}{K{_B}T}))].    
\end{multline}
Near zero temperature, the above equation can be written as
\begin{equation}
    \frac{\alpha^A_{ij}}{T} = -\frac{\pi^2}{3}\frac{K_{B}^2}{e}\frac{d\sigma\textsubscript{ij}}{d\mu}
\end{equation}
where $\alpha^A_{ij}$, $K_{B}$, $\sigma\textsubscript{ij}$ and $\mu $ are the ANC, Boltzmann constant, AHC, and chemical potential, respectively.
 
\section{RESULTS AND DISCUSSION}
\begin{figure}[t]
    \centering
    \includegraphics[width=0.5\textwidth]{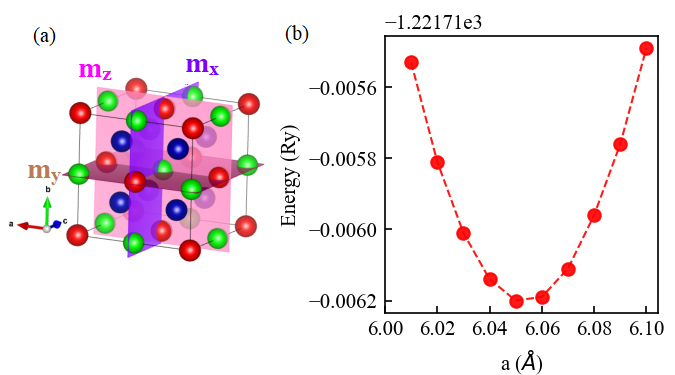}
    \caption{ (a) An unit cell of Cu$_2$CoSn Heusler compound. Blue, red, and green colors represent the Cu, Co, and Sn atoms, respectively. Three perpendicular mirror planes are designated as \textit{m$_x$}, \textit{m$_y$}, and \textit{m$_z$}, respectively.
  (b) Energy versus lattice parameter curve for the Cu$_2$CoSn Heusler compound.}
    \label{Fig1}
\end{figure}

The unit cell of  Cu$_2$CoSn full Heusler compound (space group  Fm$\bar{3}$m (No.225)) is shown in Fig.\,\ref{Fig1}(a).  The special Wyckoff's positions 8c (0.25, 0.25, 0.25), 4b (0.5, 0.5,0.5), and 4a (0, 0, 0) were considered for Cu, Co, and Sn atoms, respectively. The crystal structure has space inversion symmetry with three perpendicular relevant mirror planes. Figure \ref{Fig1}(b) shows the energy versus lattice parameter curve, which suggests the lattice parameter a = b = c = 6.05 Å for the present system. The compound is ferromagnetic with a magnetic moment of 1.15 $\mu_B$ per formula of the unit cell. The cobalt atom contributes exclusively to the magnetization ({$\mu_{Co}$} = 1.147 $\mu_B$/f.u.) as Cu and Sn are the non-magnetic elements. The non-integer magnetic moment suggests that the system deviates from the half-metallic behavior. In the absence of SOC, the crystal symmetry of magnetic Cu$_2$CoSn full Heusler compound belongs to space group Fm$\bar{3}$m, which exhibits three relevant mirror reflection symmetries \textit{m}$_x$=0, \textit{m}$_y$=0 and \textit{m}$_z$=0 in the planes \textit{k}$_x$=0, \textit{k}$_y$=0 and \textit{k}$_z$=0, respectively \cite{prb1,PhysRevB.99.165117,PhysRevB.98.241106}. In each of these planes, there is a mirror symmetry protected NL in the Brillouin zone derived from the opposite eigenvalue of mirror symmetries and cross each other at six distinct points \cite{PhysRevB.99.165117, PhysRevB.98.241106,Weyl}. These NLs gap out in the presence of SOC according to the magnetization direction, e.g., if the magnetization is considered along [001] direction, then the mirror symmetries \textit{m$_x$} and \textit{m$_y$} are no longer symmetry planes, while the \textit{m$_z$} remains the symmetry plane, as the z-component of the spin S$_z$ is left invariant by \textit{m$_z$}. Therefore, the NL in the \textit{k}$_x$ = 0 and \textit{k}$_y$ = 0 planes gap out, while the NL in the \textit{k}$_z$=0 remain still protected by the mirror reflection symmetry. The total outward Berry flux from the gapless NL is zero, while the gapped NLs produce the non-zero Berry flux in the surrounding area \cite{prb1,PhysRevB.100.054445}. The NLs in the \textit{k}$_x$ = 0 and \textit{k}$_y$ = 0 planes gapped out due to SOC result into the band anti-crossings, which restricts the Berry curvature to be aligned in magnetization direction \cite{PhysRevX.8.041045}.

\begin{figure}[t]
    \centering
    \includegraphics[width=0.52\textwidth]{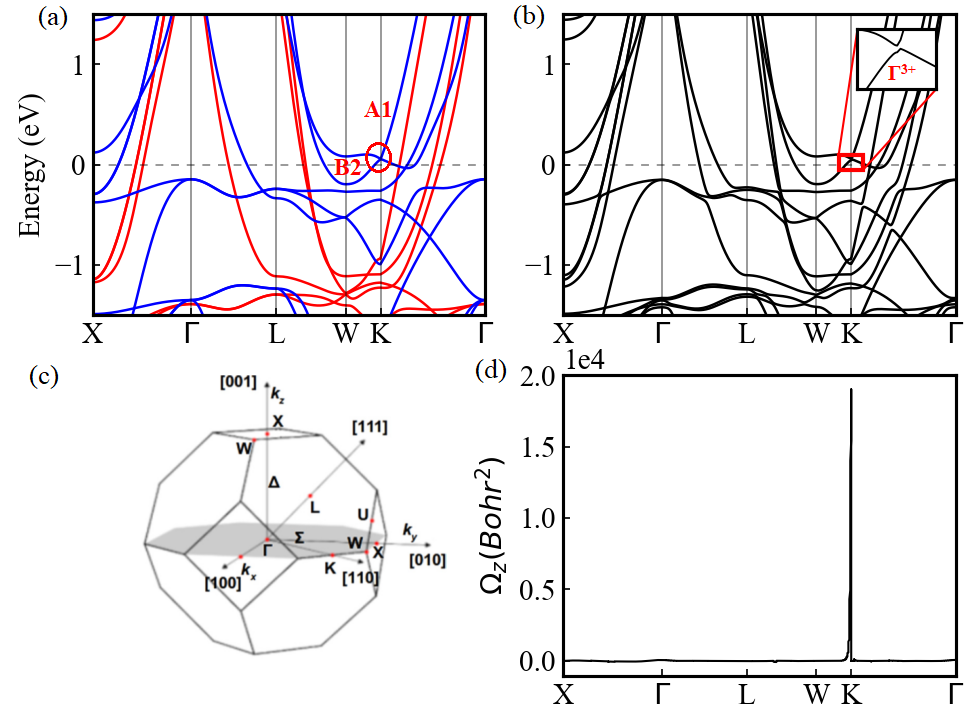}
    \caption{(a) Spin-polarized band structure (spin-up: red; spin-down: blue). (b) Band structure with spin-orbit coupling (SOC). The inset shows a zoomed view around the crossing point K. (c) The Brillouin zone in the conventional unit cell setting. (d) \textit{k}-resolved Berry curvature for the Cu$_2$CoSn.}
    \label{fig2}
\end{figure}

The spin-polarized band structure (absence of SOC) of Cu$_2$CoSn is presented in Fig.\,\ref{fig2}(a). The red and blue colors represent the majority and minority states, respectively. In the band structure of the present system, we observed an interesting linear band crossing point at high-symmetry point K close to the Fermi energy (shown inside the circle). The crossing point is made from the minority spin bands and supposes to form the NL-like band structure in \textit{k}-plane. \textbf{We did the symmetry analysis to analyze the formation of the nodal line in the system. When SOC is not considered there is no symmetry relation between the spin-up and down states and can be treated separately. The analysis of band symmetry along \textit{W}-\textit{K}-\textit{${\Gamma}$} direction, which lie on the \textit{k}$_z$=0 plane of conventional Brillouin zone of FCC lattice was done by Irrep software \cite{iraola2022irrep}. The {\lq searchcell\rq} tag was enabled for the transformation of the crystal coordinate into the cartesian coordinate.  We found two symmetry operations for the interesting crossing point at high-symmetry point K  (i) Identity (E) and (ii) two-fold rotation symmetry along [001]  direction with an inversion center. The obtained matrix operation for the band was found  
\begin{equation*}
R({\theta}) =
\begin{bmatrix}
1 & 0 & 0\\
0 & 1 & 0\\
0 & 0 &-1\\
\end{bmatrix},
\end{equation*}
which is a matrix corresponding to the \textit{m}$_z$ mirror reflection symmetry, that derives the nodal line in \textit{k}$_z$ = 0 plane. The valence and conduction bands which meet at point K [ In DFT cell \textit{i.e} crystal coordinate (0.375, 0.375, 0.750), Cartesian coordinate (-0.75, 0.75, 0) ] belong to the different irreducible groups B$_2$ and A$_1$, respectively and protected by C$_{2v}$ (mm2) point group symmetry in the space group Fm$\bar{3}$m.  Hence the crossing point formed by the intersection of  B$_2$ and A$_1$ bands form the two-fold nodal point at high-symmetry point K.}
Since SOC plays a pivotal role to realize the anomalous transport in materials and is also ubiquitous in materials with 3d elements \cite{prb1}, therefore it is necessary to study the band structure with non-vanishing SOC.
When SOC is included, we consider the hybridization of the majority and minority spin bands. The spin-up and spin-down energy bands cannot be distinguished separately because the spin no longer remains a good quantum number in the presence of SOC. The band structure in presence of SOC is shown in Fig.\,\ref{fig2}(b). The crossing point which is the interest of feature seems fragile for the SOC, where the degeneracy of the band is lost due to SOC (B$_2$ and A$_1$ transform into $ \Gamma$$^{3+}$) and a gap open between the bands (as \textit{W}-\textit{K}-\textit{${\Gamma}$} are not in the \textit{k$_z$}=0 plane in the crystal coordinate).  The inset shows the enlarged view around the crossing point.  Figure\,\ref{fig2}(c) is for the Brillouin zone of FCC lattice, which shows that  \textit{W}, \textit{K}, and \textit{${\Gamma}$} high symmetry points lie on \textit{k$_z$}=0 plane of conventional Brillouin zone. Next, we calculated the \textit{k}-resolved Berry curvature along the same high-symmetry path chosen for band structure and found that a sharp peak of Berry curvature at point K and negligible Berry curvature from the other bands (Fig.\,\ref{fig2}d), therefore the Berry curvature distribution in surrounding the Fermi surface arises from gapped nodal line greatly affect the conduction electrons and produces a large AHC and ANC in the system (discussed later).  

\begin{figure*}[t]
\centering
\includegraphics[width=1\textwidth]{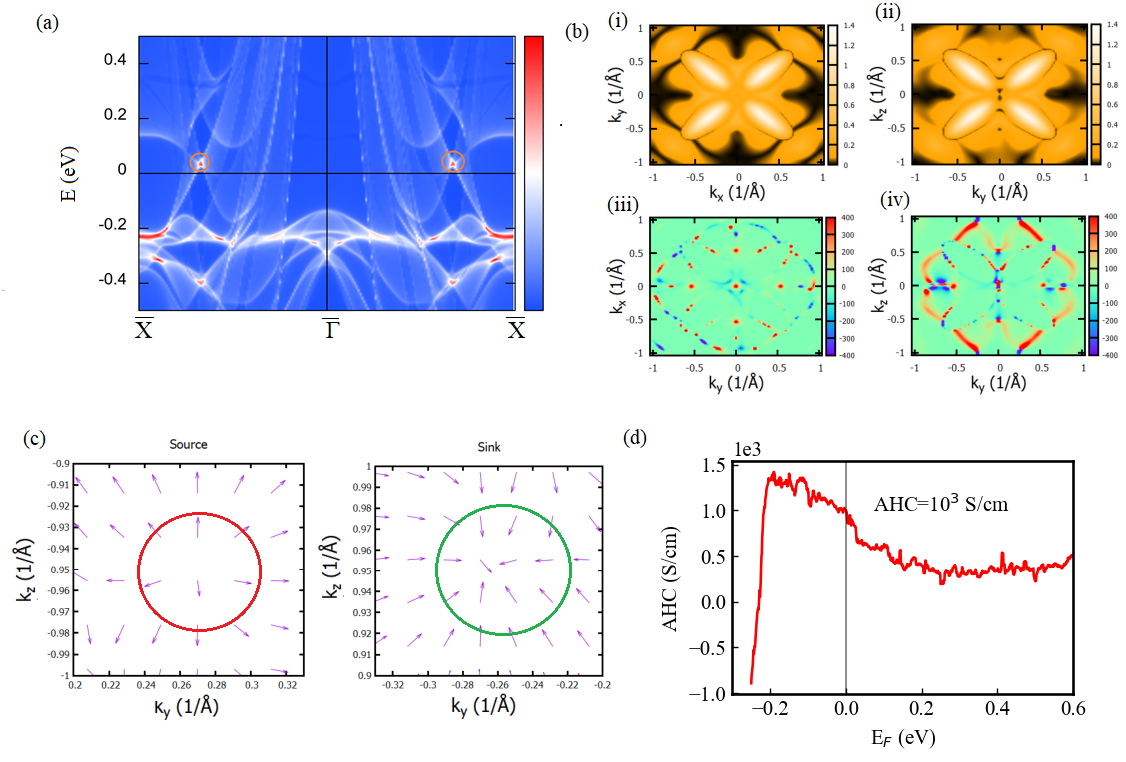}
\caption{ (a) Surface states spectrum of Cu$_2$CoSn obtained from the projection of the bulk band structure on the (001) surface. (b) The energy gap in the (i) \textit{k}$_z$ = 0 plane, (ii) \textit{k}$_x$ = 0 plane. The black color represents the vanishing gap between the bands.  Berry curvature distribution in the (iii) \textit{k}$_z$ =0 plane, (iv) \textit{k}$_x$ =0 plane. (c) The normalized Berry curvature for the Weyl points in \textit{k}$_x$ = 0 plane. The Weyl points act as the source and drain of the Berry flux. (d) The variation of AHC with Fermi energy.} 
\label{Fig3}
\end{figure*}

To inspect the topological states in our band structure, we projected the bulk band structure of Cu$_2$CoSn on the (001) surface along the  {$\overline{X}$}-{$\overline{\Gamma}$}-$\overline{X}$ direction (Fig.\,\ref{Fig3}(a)). A clear mark encircled in the surface spectrum suggests the presence of the topological band crossings, which corresponds to the linear crossing point at high-symmetry point K, and the red spot represents the gap opening at the crossing point. The small gap opening between the bands makes the denominator of Eq. (1) small and the large Berry curvature arises in the system. 
To calculate the topological properties for \textit{e.g.} Berry curvature, AHC, and ANC, etc., we constructed the MLWF from the Bloch states using Wannier90 code and found a good match between the electronic and Wannier interpolated band structure. The Wannier interpolation is an effective tool to calculate the \textit{k}-space integrals, which are involved to find out several properties of materials such as AHC, ANC, spin Hall conductivity, optical properties, etc. The MLWF method is popular to construct the Wannier functions, which is implemented in the WANNIER90 code \cite{marzari1997maximally,souza2001maximally}. In this method, the Wannier functions are generated by the unitary transformation of the Bloch wave and there is no chance of loss of information \cite{PhysRevB.105.035124}.

For a better understanding of the nature of band crossing at high-symmetry point K, we calculated the band gap in the different two-dimensions \textit{k}-planes considering the magnetization quantization axis along the [001] direction.
Figure\,\ref{Fig3}b(i) shows the energy gap in the \textit{k}$_z$ = 0 plane, which still preserves mirror symmetry. As a consequence, a closed NL is observed in this plane as shown in the black color, which is protected by the \textit{m$_z$} mirror reflection symmetry. The Berry curvature was calculated in the same \textit{k}$_z$ = 0 plane as presented in Fig.\,\ref{Fig3}b(iii), which shows that the Berry curvature around the preserved NL is very weak. 
It is interesting to look at the NL and Berry curvature in the \textit{k}$_x$ = 0 plane, which is not a plane of symmetry after considering the SOC and the magnetization direction. The NL, which was preserved in the \textit{k}$_z$ = 0 plane, gapped out in the \textit{k}$_x$ = 0 plane (Fig.\,\ref{Fig3}(b)(ii)), because of the mirror symmetry in this plane breaks upon considering the SOC and magnetization direction. The Berry curvature distribution in the same \textit{k}$_x$ = 0 plane is shown in Fig.\,\ref{Fig3}b(iv). As expected, a strong Berry curvature induces along the gapped NL, which can manifest a large transverse response in the system. A similar kind of NLs and the Berry curvature is also expected in the \textit{k$_y$} = 0 plane. Since the mirror symmetry is broken in both \textit{k$_y$} = 0 and \textit{k$_x$} = 0 planes upon considering SOC and [001] magnetization, hence the Weyl points may emerge in these planes. The mirror symmetry is still preserved in \textit{k$_z$} = 0 plane, therefore the Weyl point cannot be in the \textit{k$_z$} = 0 plane. Noteworthy, these Weyl points do not exist in the system naturally due to SOC but rather derived from the NLs, because at some \textit{k}-points the NLs refuse to break out \cite{Weyl,chang2016room}. The Berry curvature due to Weyl points derived from the gapped NL is typically small as sometimes they lie far away from the Fermi level and/or due to other Weyl points present in the same plane \cite{Weyl,chang2016room}. The energy and momentum space location of the Weyl points in possible \textit{k}-planes are mentioned in Table 1.

To further confirm the obtained points as the Weyl points, we plotted the normalized Berry curvature enclosing the coordinates of the points in \textit{k$_x$}=0 plane (Fig.\,\ref{Fig3}(c)). We found that the Weyl point of chirality + 1 acts as a source of Berry curvature (outward flux in Fig.\,\ref{Fig3}(c)) and the Weyl point with chirality -1 acts as a sink of Berry curvature (inward flux in Fig.\,\ref{Fig3}(c)). 
The strong enhancement in the Berry curvature around the gaped NLs is supposed to create the large AHC in the system. For this, we calculated the AHC by the integration of Berry curvature of all occupied dispersion bands using Eq. (1) and Eq.(2). The underlying space group with the magnetization along [001] direction contains the 4$_{001}$ symmetry operation and after summing the Berry curvature over whole Brillouin zone forces $\Omega_x$ = $\Omega_y$ = 0 and follow the relation \cite{samathrakis2022tunable}
\begin{equation}
 \begin{split}  
     -\Omega_x(k_x,k_y,k_z) = \Omega_x(-k_x,-k_y,k_z)\\
     -\Omega_y(k_x,k_y,k_z) = \Omega_y(-k_x,-k_y,k_z)\\
      \Omega_z(k_x,k_y,k_z) = \Omega_z(-k_x,-k_y,k_z).
\end{split}
\end{equation}
Therefore, following the symmetry operation the z-component of AHC $\sigma^A_z$ is unrestricted, while $\sigma^A_x$ and $\sigma^A_y$ identically vanish.
The variation of AHC with Fermi energy is shown in Fig.\,\ref{Fig3} (d). We found the giant intrinsic AHC ($\sigma^A_z$) about 1003 \textit{S/cm} at the Fermi energy, which varies to 1120\textit{S/cm} just 0.05 eV below the Fermi level. This magnitude of AHC is larger than most of the investigated systems \cite{prb1,chen2022large, mende2021large,asaba2021colossal,chen2021anomalous,wang2017anisotropic} and comparable to the highest AHC reported for Co$_2$MnGa Heusler compound \cite{guin2019anomalous, sakai2018giant}.
  \begin{table}[htbp]
  \centering
   \begin{tabular}{lrrrrr}
   \midrule\midrule
   Weyl point & \multicolumn{1}{l}\,\,{$k_x$ (2$\pi$/a)} & \multicolumn{1}{l}\,\,{$k_y$ (2$\pi$/a)} & 
   \multicolumn{1}{l}\,\,{$k_z$(2$\pi$/a)} & \multicolumn{1}{l}{         \,\,E (eV)} &  \\
   \midrule
     W$_{\pm{A}}$    & \,0.26     \,\,\,& 0.00  & \,\,\,$\pm{0.95}$  & -0.240  &  \\
    W$_{\pm{B}}$    & 0.00     &\,\,\, $\pm{0.26}$  &\,\,\, $\mp{0.95}$ & 0.235   &  \\
    W$_{\pm{C}}$    & 0.00     & 0.00     &   $\pm{0.51}$  & 0.73  &  \\
   \midrule\midrule
   \end{tabular}%
  \caption{Representative coordinates of Weyl points in different planes in the momentum space with their chemical potentials with reference to the Fermi energy.}
  \label{tab:addlabel}%
\end{table}%
\begin{figure}[h]
\centering
\includegraphics[width=0.4\textwidth]{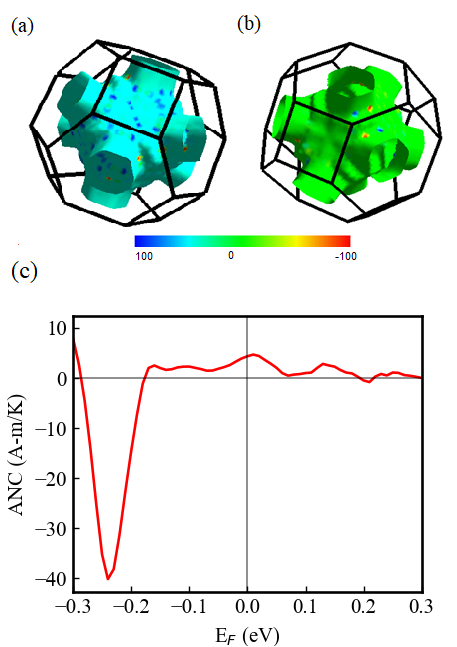}
\caption{Fermi surface and top of that (a) z-component of Berry curvature $\Omega^z$ (b) y-component of Berry curvature $\Omega^y$. (c) Fermi-level variation of anomalous Nernst conductivity.} 
\label{Fig4}
\end{figure}

 Now we discuss the ANE in the present compound. The ANE is the thermometric counterpart of AHE, where the temperature gradient is used for the motion of charges instead of the electric field \cite{PhysRevResearch.4.013215}. The origin of ANE is closely related to the AHE and the key difference is that the AHE is the summation of the Berry curvature over all occupied states, while the ANE is the sum of the Berry curvature of states close to the Fermi energy \textit{i.e} the ANC is the sum of the Berry curvature on the Fermi surface \cite{PhysRevB.98.241106,PhysRevResearch.4.013215}. The magnitude of ANC is related to the variation of the AHC near the Fermi energy. We plotted the z-component of Berry curvature ($\Omega^z$) and y-component of Berry curvature ($\Omega^y$) on the Fermi surface of the Cu$_2$CoSn system is shown in Fig.\,\ref{Fig4}(a)-(b) and using Eq.\,(3) and (4) the ANC was calculated. We found the strong $\Omega^z$ on the Fermi surface, while the $\Omega^y$ identically cancels out due to the presence of the equal amount of positive and negative hotspot of the Berry curvature on the Fermi surface as shown in Fig.\,\ref{Fig4}(b).  This strong Berry curvature on the Fermi surface gives the $\frac{\alpha^A}{T}$ = 0.013 \textit{A/m-K$^2$} and the ANC reaches to the $\sim$ 3.98 A/m-K at 300\, K, which is similar to the highest reported value of ANC is 4.0 \textit{A/m-K} in the Co$_2$MnGa experimentally. The variation of the ANC with the Fermi energy is shown in Fig.\,\ref{Fig4} (c), which shows a sudden increase in the ANC below -0.25 eV that might be related to the presence of flat band at this energy level in the band structure.
  \section{CONCLUSION}
In summary, we theoretically investigated the electronic, magnetic, and anomalous transport properties of Cu$_2$CoSn full Heusler compound. We found three NLs in the present compound, which are preserved by the mirror reflection symmetries of the system. Upon considering the SOC, the NLs gap out according to magnetization direction, consequently a strong Berry curvature develops along the gapped NL, which leads to the high AHC and ANC in the Cu$_2$CoSn Heusler compound. Therefore, the Cu$_2$CoSn is added as a new candidate in the family of Heusler compounds with high AHC and ANC. Our work provides a comprehensive understanding of the anomalous transport properties in the magnetic NL materials, specifically in the full Heusler compound, in context to the breaking of the protected mirror symmetries.
\section*{ACKNOWLEDGMENT}
We gratefully acknowledge the PARAMSHIVAY Supercomputing Centre of IIT(BHU) for cluster support. S.S. thanks the Science and Engineering Research Board
of India for financial support through the “CRG” scheme (Grant No. CRG/2021/003256) and Ramanujan Fellowship (Grant No. SB/S2/RJN-015/2017) and UGC-DAE CSR, Indore, for financial support through the “CRS” scheme. G.K.S. thanks the DST-INSPIRE scheme for support through a fellowship.

*ssingh.mst@itbhu.ac.in 

 \end{document}